\documentclass{evn2004}
\usepackage{txfonts}
\begin{document}
   \title{Current Results from the RRFID Kinematic Survey:
Apparent Speeds from the First Five Years of Data}

   \author{B.~G. Piner\inst{1}, A.~L. Fey\inst{2},
          \and
          M. Mahmud\inst{1}
          }

   \institute{Department of Physics and Astronomy, Whittier College,
13406 E. Philadelphia Street, Whittier, CA 90608, U.S.A.
         \and
U.S. Naval Observatory, 3450 Massachusetts Ave.,
Washington D.C. 20392, U.S.A.
             }

   \abstract{
We present current results from our ongoing project to study the parsec-scale
relativistic jet kinematics of sources in the U.S. Naval Observatory's
Radio Reference Frame Image Database (RRFID).  The RRFID consists
of snapshot observations using the VLBA plus up to 9 
additional antennas at 8 and 2 GHz, and is intended to allow
monitoring of these sources for variability or structural
changes so they can be evaluated for continued suitability as radio
reference frame objects.  The Image Database currently contains about
3000 images of 450 sources from 1994 to 2004, with some sources having
images at 20 epochs or more.
                                                                                                                      
We have now completed analysis of the 8 GHz images for all sources observed
at 3 or more epochs from 1994 to 1998.  The completed analysis comprises
966 images of 87 sources, or an average of
11 epochs per source.  Apparent jet speeds have been measured for these
sources, and the resulting speed distribution has been compared with results
obtained by other large VLBI surveys.
The measured apparent speed distribution agrees with those found by the 2~cm survey
and Caltech-Jodrell Bank (CJ) survey; however, when a source-by-source comparison is done with the 2~cm survey
results, significant disagreement is found in the apparent speed measurements for a number of sources.
This disagreement can be traced in most cases to either an insufficient time baseline
for the current RRFID results, or to apparent component mis-identification in the 2~cm survey results
caused by insufficient time sampling.  These results emphasize the need for long time baselines
and dense time sampling for multi-epoch monitoring of relativistic jets.
   }

   \titlerunning{Current Results from the RRFID Kinematic Survey}
   \maketitle

\section{The Radio Reference Frame Image Database}
The U.S. Naval Observatory (USNO) maintains a multi-epoch database of VLBI images 
of objects that comprise the Radio Reference Frame.  
This Radio Reference Frame Image Database (RRFID) consists of Very Long Baseline Array (VLBA) 
snapshot observations at 8 and 2 GHz (with the addition of up to 9 geodetic antennas ---
including European antennas at Medicina, Ny~Alesund, Onsala, and Wettzell),
and is intended to allow monitoring of these sources for variability or structural changes 
so they can be evaluated for continued suitability as radio reference frame objects.  
The Image Database currently contains about 3000 images of 450 sources from 1994 to 2004. 
(although the number of fully reduced epochs after 1998 is currently sparse).
The web-based interface to the RRFID is available at rorf.usno.navy.mil/RRFID/.  
The majority of objects in this database are radio-loud quasars or BL Lac objects.
A sample image of the well-known object BL Lac at epoch 1998 December 21 
(the final epoch analyzed for this paper) is shown in Figure~1,
to show the typical noise levels and beam sizes of these snapshot images. 
The RRFID has recently begun adding high-frequency images 
at 24 and 43 GHz as well.  Some of the RRFID epochs are described by 
Fey, Clegg, \& Fomalont (1996), Fey and Charlot (1997), and Fey and Charlot (2000).

\begin{figure}
   \centering
   \vspace{270pt}
   \includegraphics{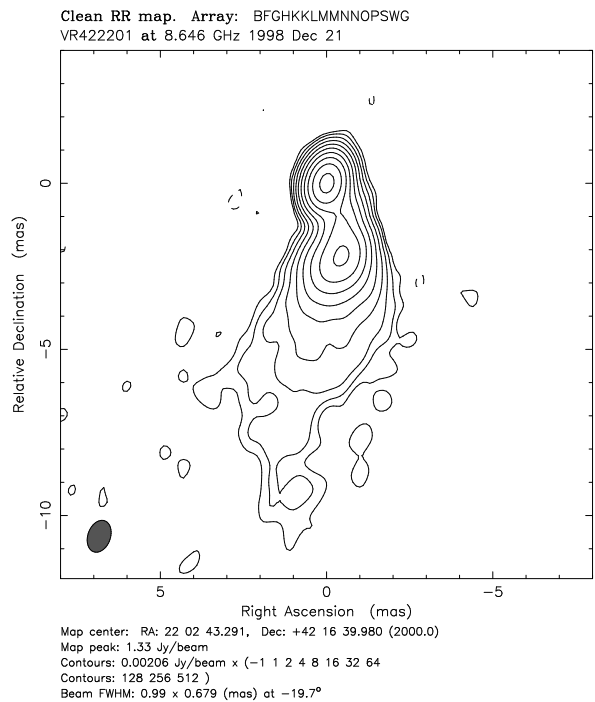}
      \caption{A sample image from the RRFID: BL Lac on 1998 Dec 21.}
   \end{figure}

\section{The RRFID Kinematic Survey}
\label{rrfidks}
Large multi-epoch VLBI surveys containing many sources are important for characterizing 
the physics of the relativistic jets.  
Measurement of the apparent speed distribution can shed light on the true distribution 
of jet Lorentz factors, and the maximum speeds present in the distribution can 
reveal the maximum kinetic energy that must be able to be supplied by the central engine.  
The apparent speed distribution can also reveal whether the motions seen in VLBI images 
are true bulk motions of the plasma, or only a pattern speed superimposed on the underlying jet.  
Large VLBI surveys should also yield interesting results in studies of jet bending.  
Some individual jets have been reported to follow helical trajectories (e.g., Hong et al. 2004), 
possibly due to precession of the jet nozzle induced by torques in binary supermassive black hole systems.  
A large survey of hundreds of parsec-scale jets enables the question of helical jets to be studied statistically.

We are currently engaged in a project to extract the information on jet 
kinematics and dynamics that is contained in the RRFID.  Our project consists of two major parts:
\begin{enumerate}
\item{A multi-epoch kinematic survey of the 8 GHz images of all sources 
that have been observed at 3 or more epochs, 
for the purpose of measuring apparent jet velocities from the motions of VLBI components.}
\item{An analysis of single-epoch images at 8 and 2 GHz for all 450 sources 
for the purposes of obtaining those measurements that can be made from single-epoch images, 
such as: bending of the parsec-scale jet ridgeline, 
misalignment with kiloparsec-scale structure, apparent jet opening angles, and core and component spectral indices.} 
\end{enumerate}
Some RRFID sources are observed quite frequently, 
the best-observed sources included in this paper have been observed at 19 epochs through the end of 1998.  
We are currently working on the multi-epoch portion of this project (the kinematic survey), 
and have completed the analysis of all sources observed at 3 or more epochs from 1994 through 1998.
This list comprises 87 sources, with the average source being observed at 11 epochs, so that
a total of 966 images and model fits have been analyzed for this paper. 
Note that over 30 epochs that will eventually be added to the RRFID
have also been observed since the end of 1998, but most of this
more recent data has yet to be reduced.

The current results are comparable in terms of the number of images to the other
two large VLBI surveys: the NRAO 2~cm survey (Kellermann et al. 2004), and the Caltech-Jodrell Bank (CJ)
survey (Britzen et al. 2001; Vermeulen 1995).  
However (in particular when compared to the CJ survey), we
have a smaller number of sources with a much higher number of epochs per source, and therefore
much denser time sampling. It is also important to note that the RRFID is {\em not} a flux-limited
sample.

\section{Apparent Speeds}
The apparent speed measurements were made as follows.  
The self-calibrated visibility data corresponding to the CLEAN images on the RRFID 
web site were loaded into the DIFMAP software package, 
and circular Gaussian components were fit to the visibilities.  
Occasionally an elliptical Gaussian component was used to represent the core 
or an especially bright or well-resolved jet component.  
The distances of the Gaussian components from the presumed
core were recorded and components were identified from epoch to epoch.  
Positional error bars were assigned as a fraction of the beam size 
that decreased with increasing component surface brightness and the presence or absence of other confusing components.  
A linear least-squares fit was then done to the separation from the core versus time to yield a proper motion.  
The proper motion was converted to an apparent speed using the following 
cosmological parameters: Ω$\Omega_{m}=0.27$, $\Omega_{\Lambda}=0.73$, and $h=0.71$.  
Two sample fits are shown in Figure~2, for the two sources 0552+398 and 1308+326.  
The component separations for these two sources show a very small scatter about the linear fits
(only a few percent of the beam width), demonstrating the ability to make precise positional
measurements with this snapshot data.
The plot for 0552+398 shows the common phenomenon of a stationary component, presumably
a stationary pattern on the underlying relativistic jet.
The plot for 1308+326 shows one of our highest confidence detections of a fast apparent speed;
the measured apparent speed is $18~c$.

\begin{figure}
   \centering
   \vspace{400pt}
   \includegraphics{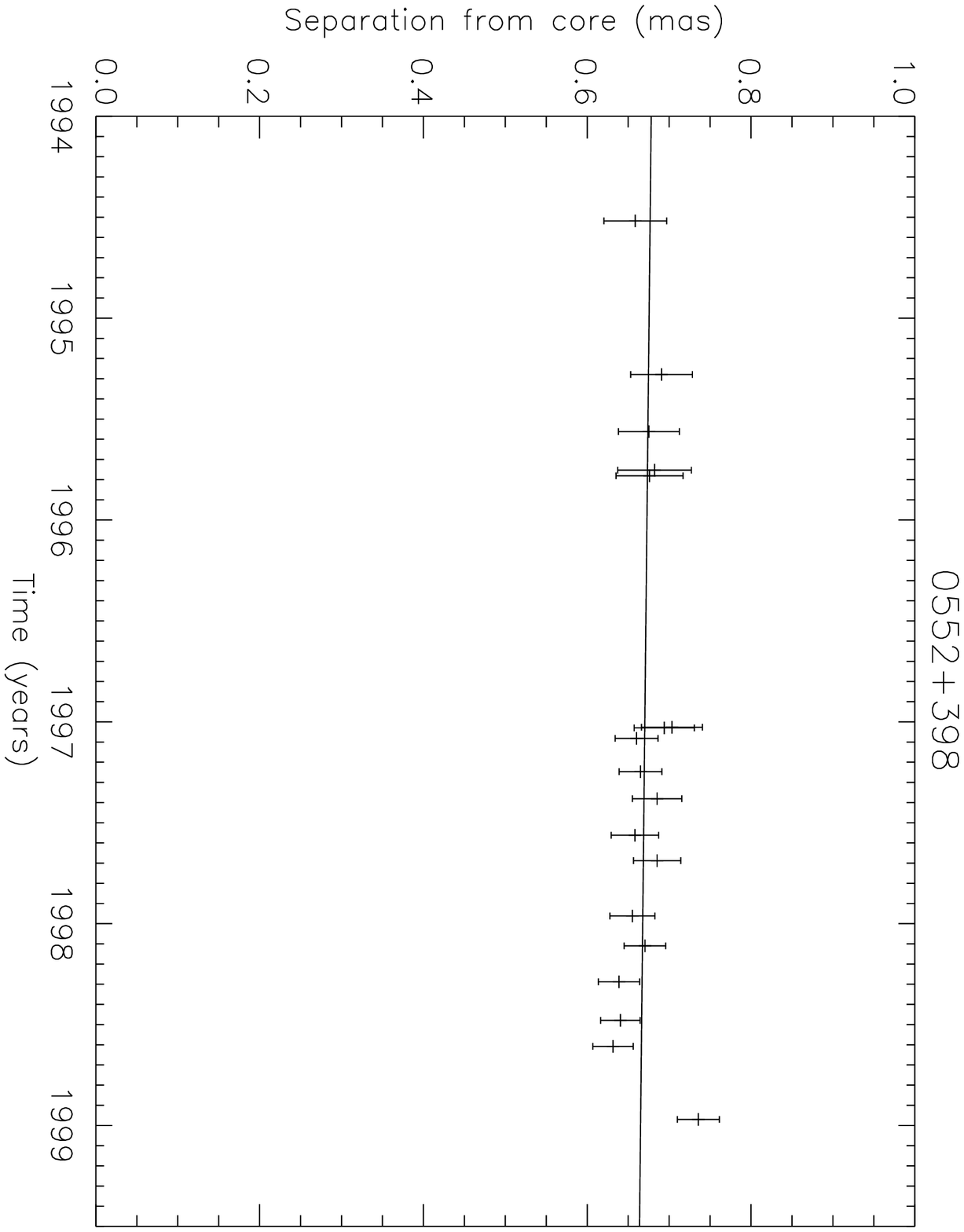}
   \includegraphics{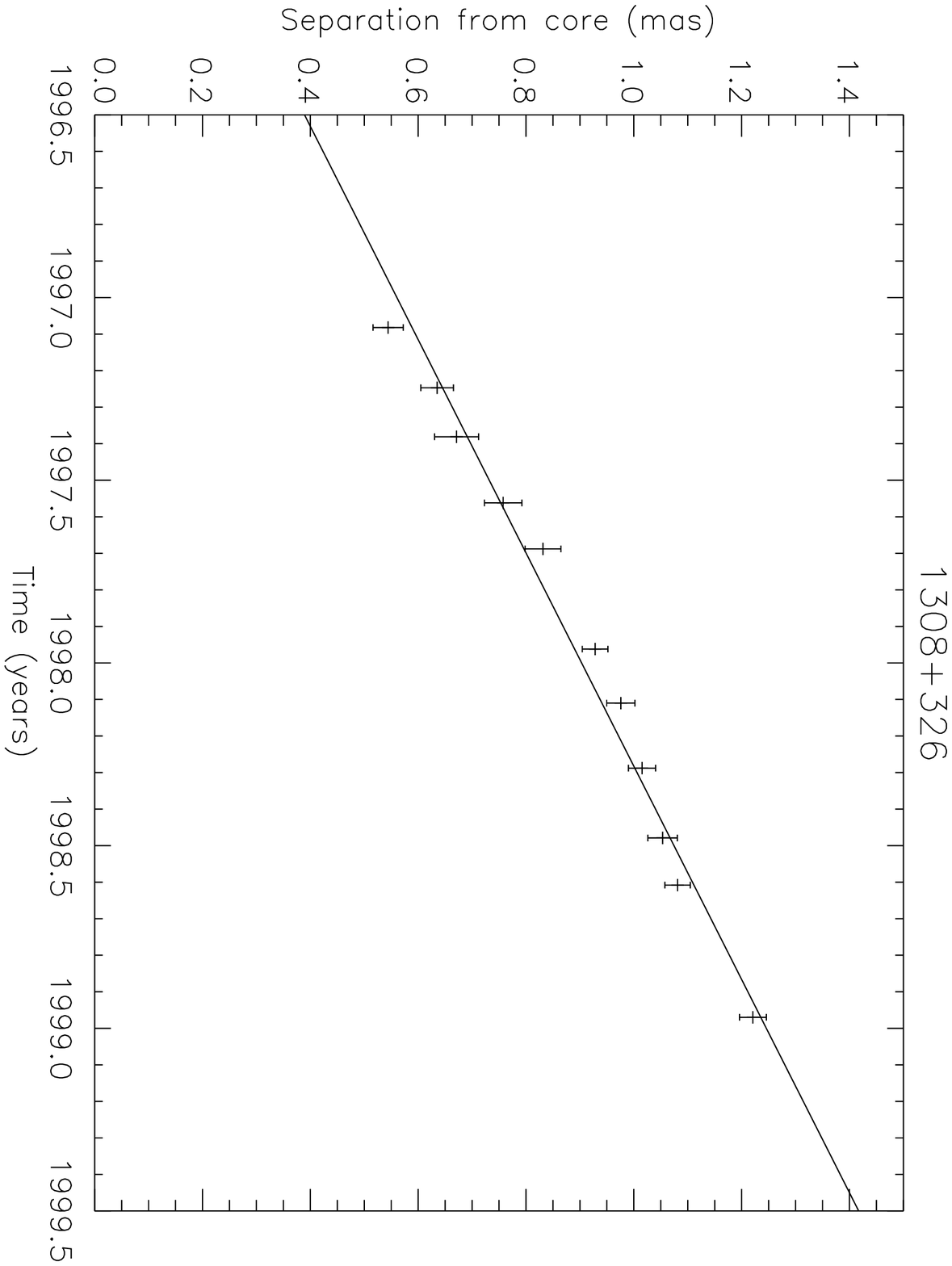}
   \caption{Fits to component separations vs. time for 0552+398 and 1308+326.}
   \end{figure}

The histogram of apparent speeds for all components with apparent speed errors $<10~c$ 
is shown in Figure~3 (140 components,
some components had large errors associated with their apparent speed measurement because they were
not added to the observing sessions until the last year of the data considered in this paper). 
The histogram has a peak at slow apparent speeds, 
and a tail extending out to $30~c$.  The mean apparent speed for all 140 components is $4.5~c$.  
The high-speed tail of the distribution indicates that Lorentz factors up to 
$\approx 30$ (and true speeds up to $0.999~c$) 
exist in some sources, and must be able to be produced by theories for jet acceleration.  
If the quasars in Figure~3 are separated from the BL Lac objects, then a KS test detects
a difference in the apparent speeds of the two groups at low significance (92\%).
The mean apparent speed of the quasar components is $5.3~c$, that of the BL Lac components
is $2.3~c$.

\begin{figure}
   \centering
   \vspace{200pt}
   \includegraphics{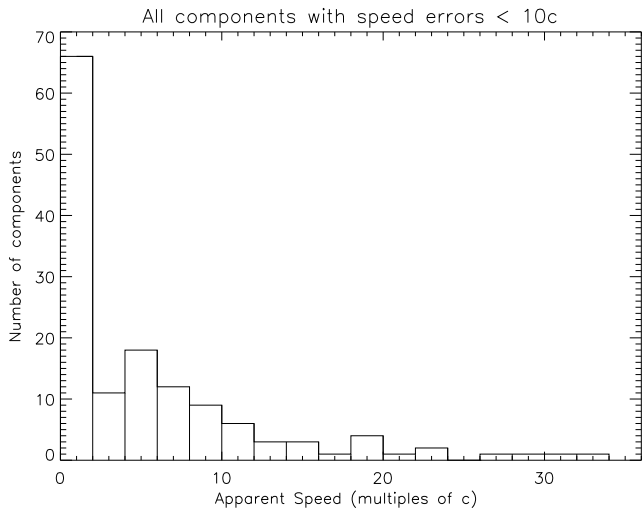}
   \caption{Histogram of measured apparent speeds of jet components.}
   \end{figure}

A KS test comparing the histogram in Figure~3 to the equivalent histograms produced by the 
2~cm survey and the CJ survey shows with high significance that the 
2~cm survey, the CJ survey, and the RRFID survey are measuring the same speed distribution 
(but since many sources overlap between these surveys, 
the KS test would only be expected to show a 
difference if speeds were changing in a systematic way with observing frequency, 
i.e. with distance from the core).  However, a KS test does show a difference at the 99.95\% 
confidence level with the histogram of apparent speeds of EGRET blazars by Jorstad et al. (2001), 
as expected if their gamma-ray selected sample is more highly beamed than our radio selected sample.  
It has been noted by the 2~cm survey authors (Kellermann et al. 2004) 
that a speed distribution like that in Figure~3 cannot be explained by the simplest model 
in which all sources have the same Lorentz factor.  
Either a power law distribution of bulk Lorentz factors, 
or separate bulk and pattern velocities, 
needs to be invoked to explain such a distribution (Kellermann et al. 2004). 

\section{Detailed Comparison with 2~cm Survey Results}
Of the 87 sources included in this paper, 42 are also members of the 2~cm survey.
In the previous section, we showed that the overall apparent speed distributions 
as measured by these two surveys agree with each other with high significance.
In this section, we investigate whether this agreement also extends to measurements
of individual components, rather than just the overall distribution.
From the 42 sources in common between this paper and the 2~cm survey, we identify
64 components that can (at least at a single epoch) be identified with a corresponding
component from the other survey.
The proper motions measured for these 64 components from both the 2~cm survey
and the RRFID survey are compared in Figure~4.
The Pearson correlation coefficient shows significant correlation between the measured proper motions;
however, this is due entirely to the four points in the upper right of the figure,
all of which are components of BL Lac.  If the components of BL Lac are removed from the comparison,
then there is no significant correlation.

\begin{figure}
   \centering
   \vspace{200pt}
   \includegraphics{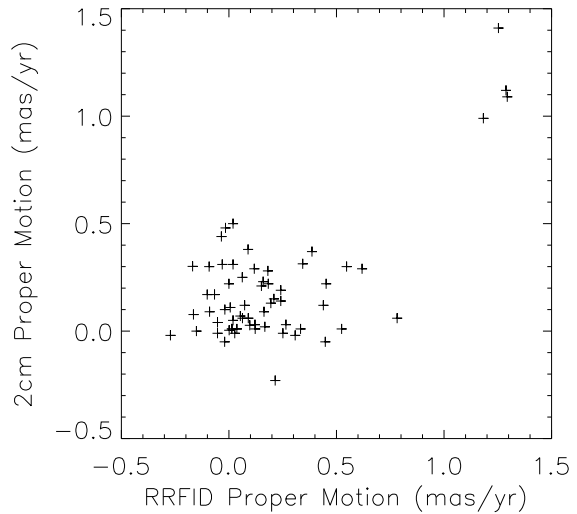}
      \caption{Comparison of proper motions measured in the 2~cm survey
and RRFID survey for common components.}
   \end{figure}

For many of the common components there is apparently significant disagreement 
in the measured proper motion between these two surveys,
and this is what causes the scatter plot
appearance of Figure~4. 
This is a cause for significant concern
for studies that rely on the robustness of VLBI model-fitting results on specific sources.
In order to determine the causes of the disagreements in the proper motion measurements, we have made
a detailed comparison of the proper motion results from these two surveys
by over plotting the 2~cm survey component separations and linear fits on top
of our results.
For approximately half of the common components, the measured proper motions from
the two surveys agree within their stated errors, and for a number of these the
fitted motions lie virtually on top of each other.
For the other half, we have made a component-by-component examination of the cause of the
differences in the proper motion measurements.
In the majority of cases where the two surveys have observations at nearly simultaneous epochs,
the {\em fitted separations} of model components between the two surveys {\em agree},
despite the differences in observing frequency and fitting algorithm used.
The disagreement in the proper motions in some cases can be traced to the 
differences in the {\em identification} of components 
from epoch to epoch, in others to an insufficient time baseline for the RRFID observations,
which are currently reduced only through 1998.

Examples of these two sources of disagreement are shown in Figure~5.
Points from the RRFID are shown with error bars and with solid-line fits, points from the
2~cm survey are shown without error bars and with dotted-line fits.
In the top panel note that the disagreement in the proper motion of the inner component
is due to the extra 2~cm survey point in 2001; disagreements of this type
should be resolved once the unreduced epochs in the RRFID are analyzed. 
The bottom panel shows a case where the discrepancy could be resolved by re-labeling the
components from the 2~cm survey to match our identifications, i.e., a fewer number of slower
components instead of a greater number of faster components. 
In particular, note that if one follows along the solid line representing the fit to the second component in the
RRFID results (the diamonds), you encounter successively what has been identified as three faster components
in the 2~cm survey: represented by diamonds, triangles, and squares.
In this case the dense time sampling of the RRFID results in 1997 and 1998 seems to preclude
the faster speeds measured by the 2~cm survey.
It seems that for apparent jet speeds derived from VLBI model fitting to be truly robust
requires {\em both} a long time baseline {\em and} dense time sampling.

\begin{figure}
   \centering
   \vspace{400pt}
   \includegraphics{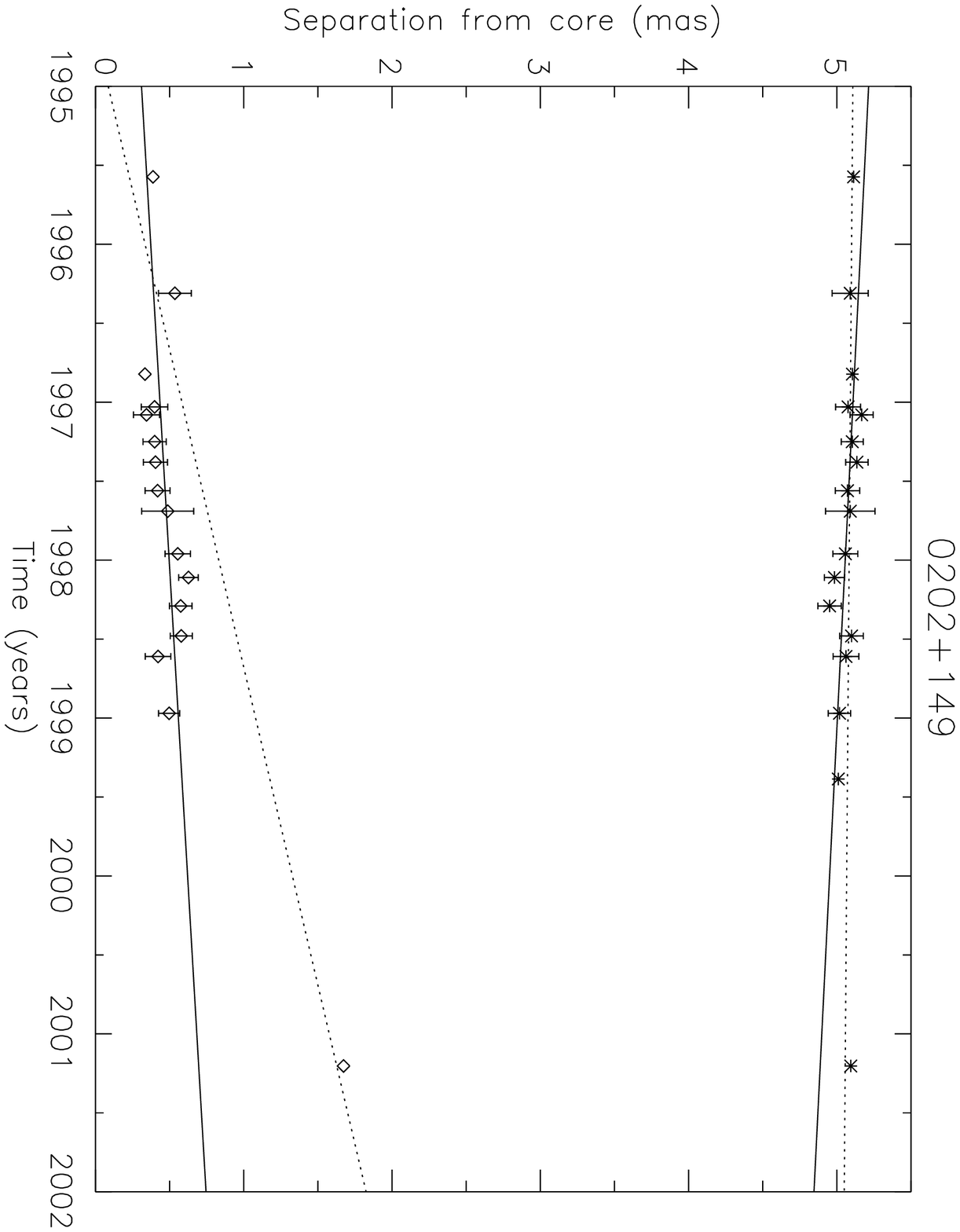}
   \includegraphics{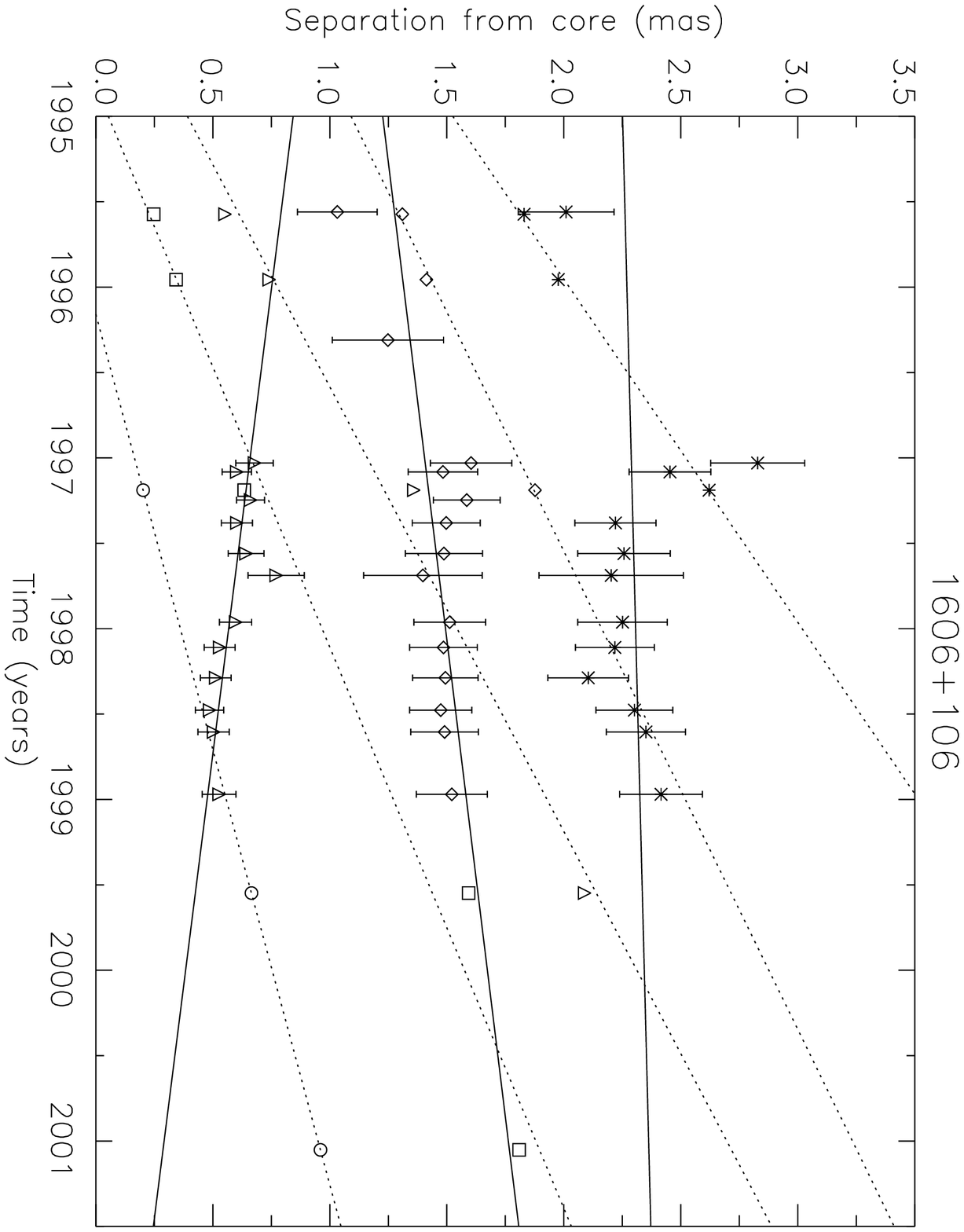}
   \caption{Fits to component separations vs. time for 0202+149 and 1606+106.
The points with associated error bars and the solid-line fits represent data from the RRFID survey.
The points with no error bars and the dotted-line fits are the results of the 2~cm survey
(Kellermann et al. 2004).  Components are identified by asterisks, diamonds, triangles, squares,
and circles from the outermost identified component inward.}
   \end{figure}

\section{Conclusions}
Although about 50 epochs have been observed for the RRFID program over the last ten years, only about the
first 20 epochs from 1994-1998 have been fully reduced. 
If the additional 30 epochs obtained from 1999-2004 are eventually added to those already reduced for this
paper, then the RRFID Kinematic Survey will be by far the largest survey of jets
(in terms of number of images) yet made, and will have the advantages of both a long time baseline (10 years),
and dense time sampling ($\approx 5$ epochs/year).  This survey also has the obvious advantage that the
VLBI observations comprising it have already been made, so it remains only
to reduce the data.

Meanwhile, work is continuing on the current data set, including searches for non-linear
motions of jet components, and analysis of parsec to kiloparsec misalignment angles.

\begin{acknowledgements}
The National Radio Astronomy Observatory is a facility of the National Science Foundation operated
under cooperative agreement by Associated Universities, Inc.
This research has made use of the United States Naval Observatory (USNO) Radio Reference Frame Image Database (RRFID).
We thank Ken Kellermann for providing model component data from the 2~cm survey.
This work was supported by the
National Science Foundation under Grant No. 0305475,
and by a Cottrell College Science Award from Research Corporation.
\end{acknowledgements}

\end{document}